\documentstyle[12pt,aasms4]{article}

\received{24 July 1998}
\accepted{27 August 1998}

\lefthead{Lamarre et al.}
\righthead{First measurement of the submillimetre S.Z. effect}

\begin{document}

\newcommand{\dix}{\,{ \,10} }
 \newcommand{\mic}{\,{ \mu m} }
 \newcommand{\tiretmic}{\,{- \mu m} }
 \newcommand{\pac}{\,{ pc} }
 \newcommand{\Wmsr}{\,{ Wm^{-2}sr^{-1}} }
 \newcommand{\Wmsrmic}{\,{ Wm^{-2}sr^{-1}\mic^{-1}} }
 \newcommand{\Wmsrhz}{\,{ Wm^{-2}sr^{-1}Hz^{-1}} }
 \newcommand{\Wcmsr}{\,{ Wcm^{-2}sr^{-1}} }
 \newcommand{\Wcm}{\,{ Wcm^{-2}} }
 \newcommand{\cm}{\,{ cm^{-2}} }
 \newcommand{\cmmun}{\,{ cm^{-1}} }
 \newcommand{\Wm}{\,{ Wm^{-2}} }
 \newcommand{\WH}{\,{ W/H_{atom}} }
 \newcommand{\Wcmmic}{\,{ Wcm^{-2}\mic^{-1}} }
 \newcommand{\Wmmic}{\,{ W/m^2/\mic} }
 \newcommand{\MJysr}{\,{ MJysr^{-1}} } 
 \newcommand{\Wcmmicsr}{\,{ W/cm^2/\mic /sr} }
 \newcommand{\Dl}{\,{ \Delta \lamb} }
 \newcommand{\lamb}{\,{ \lambda} }
\newcommand{\Inu}{\,{ I_{\nu}}}
 \newcommand{\nuInu}{\,{ \nu I_{\nu}}}
 \newcommand{\nHdeux}{\,{n_{H_2}} }
\newcommand{\nHtwo}{\,{n_{H_2}} }
\newcommand{\NHdeux}{\,{N_{H_2}} }
\newcommand{\nHII}{\,{n_{HII}} }
\newcommand{\nH}{\,{n_{H}} }
\newcommand{\NH}{\,{N_{H}} }
\newcommand{\cmcube}{\,{ cm^{-3}} }
\newcommand{\cmdeux}{\,{ cm^{-2}} }
\newcommand{\cmdeuxkms}{\,{ cm^{-2}(km/s)^{-1}} }
\newcommand{\cmun}{\,{ cm^{-1}} }
\newcommand{\cmcubes}{\,{ cm^{-3}s^{-1}} }
\newcommand{\kcmcubes}{\,{ cm^{3}s^{-1}} }
\newcommand{\tUV}{\,{\tau_{\mbox{UV}}} }
\newcommand{\tV}{\,{\tau_{\mbox{V}}} }
\newcommand{\Lsol}{\,{ L_{\sun}} }
\newcommand{\IO}{\,{ I_0} }
\newcommand{\IL}{\,{ I_L} }
\newcommand{\epstrois}{\, {\varepsilon_{3.3}} }
\newcommand{\Hbeta}{\,{ H_{\beta}} }
\newcommand{\Halpha}{\,{ H_{\alpha}} }
\newcommand{\Bralpha}{\,{ Br_{\alpha}} }
\newcommand{\pccmsix}{\,{ pc \, cm^{-6}} }
\newcommand{\Htwo}{\,{ H_2} }
\newcommand{\CeighteenO}{\,{C^{18}O} }
\newcommand{\Av}{\,{ A_V} }
\newcommand{\Cplus}{\,{ C^+} }
\newcommand{\Etrois}{\, {E_{3.3}} }
\newcommand{\FC}{\, {F_{\lambda}(C)} }
\newcommand{\FK}{\, {F_{\lambda}(K)} }
\newcommand{\FL}{\, {F_{\lambda}(L)} }
\newcommand{\FPAH}{\, {F_{\lambda}(PAH)} }
\newcommand{\hnu}{\, { h\nu}}
\newcommand{\douzeCO}{\, { ^{12}CO}}
\newcommand{\treizeCO}{\, { ^{13}CO}}
\newcommand{\Hplus}{\, { H^+}}
\newcommand{\HdeuxO}{\,{ H_{2}0}}
\newcommand{\NHtrois}{\,{ NH_{3}}}
\newcommand{\orthoNHtrois}{\,{ ortho-NH_{3}}}
\newcommand{\paraNHtrois}{\,{ para-NH_{3}}}
\newcommand{\HtwoeighteenO}{\,{ H_{2}^{18}O}}
\newcommand{\HtwoO}{\,{ H_{2}O}}
\newcommand{\orthoHtwoO}{\,{ortho-H_{2}O}}
\newcommand{\paraHtwoO}{\,{para-H_{2}O}}
\newcommand{\Odeux}{\,{ O_{2}}}
\newcommand{\Otwo}{\,{ O_{2}}}
\newcommand{\Fd}{\,{ F_d}}
\newcommand{\betal}{\,{ \beta_{l}}}
\newcommand{\betad}{\,{ \beta_{d}}}
\newcommand{\gu}{\,{ g_{u}}}
\newcommand{\gl}{\,{ g_{l}}}
\newcommand{\Eu}{\,{ E_{u}}}
\newcommand{\El}{\,{ E_{l}}}
\newcommand{\Aul}{\,{ A_{ul}}}
\newcommand{\nup}{\,{ n_{u}}}
\newcommand{\smun}{\,{ s^{-1}}}
\newcommand{\Ho}{\,{H_0}}

\title{First measurement of the submillimetre  \\
Sunyaev-Zel'dovich effect}

\author{Lamarre J.M. \altaffilmark{1},
 Giard M. \altaffilmark{2}, 
 Pointecouteau E. \altaffilmark{2},
 Bernard J.P. \altaffilmark{1}, 
 Serra G. \altaffilmark{2},
 Pajot F. \altaffilmark{1},
 D\'esert F.X. \altaffilmark{1}, 
 Ristorcelli I. \altaffilmark{2},
 Torre J.P. \altaffilmark{3}, 
 Church S. \altaffilmark{4}, 
 Coron N. \altaffilmark{1}, 
 Puget J.L. \altaffilmark{1}, 
 Bock J.J. \altaffilmark{5}}


\altaffiltext{1}{Institut d'Astrophysique Spatiale,  
B\^at 121, Universit\'{e} Paris-Sud, F-91405 Orsay cedex, France}

\altaffiltext{2}{Centre d'Etude Spatiale des Rayonnements, BP4346, F-31028 Toulouse cedex 4, France}

\altaffiltext{3}{Service d'A\'{e}ronomie du CNRS, BP3, F-91371 Verri\`{e}res le Buisson cedex, France}

\altaffiltext{4}{Department of Physics, Math, and Astronomy, California Institute of Technology, CA 91125, USA}

\altaffiltext{5}{Jet Propulsion Laboratory, 
MS 169-327, Pasadena, CA 91109, USA}


\begin{abstract}

We report the first detection of the Sunyaev-Zel'dovich 
increment on the cosmic microwave background (CMB) at submillimetre 
wavelengths in the direction of a cluster of galaxies. 
It was achieved towards 
the rich cluster Abell 2163,
using the PRONAOS 2-metre stratospheric telescope. 
Together with data from the SuZIE, Diabolo, and ISO-PHT experiments, 
these measurements, for the first time, give a complete picture of 
the far-infrared to millimetre spectral energy 
distribution of the diffuse emission toward a cluster of galaxies. 
It clearly shows the positive and negative 
parts of the S.Z. effect 
and also a positive signal at short wavelengths that can be 
attributed to foreground dust in our galaxy.

\end{abstract}


\keywords{balloons --- cosmic microwave background ---
dust, extinction --- galaxies: clusters: individual (A 2163) ---
 infrared: general}

\section{Introduction}

The Sunyaev-Zel'dovich effect (Sunyaev and Zel'dovitch \cite{Sunyaev72}, 
S.Z. in the following) is a spectral distortion 
of the Cosmic Microwave Background (CMB) due to inverse Compton scattering 
of CMB photons by hot electrons in clusters of galaxies. 
It consists of a flux decrement at millimetre
and centimetre wavelengths and a flux increment at shorter wavelengths. 
If the temperature of the electron
gas is determined independently, for instance from the Xray spectrum, then the 
knowledge of both parts of the S.Z. 
effect can provide information about the intracluster gas mass (thermal
effect), and on its
peculiar velocity with respect to the Hubble flow (kinetic effect).
Recent calculations which fully take into account the relativistic
effects, have shown that for the hottest clusters, $T_e > 8 keV$,
 the S.Z. spectrum significantly departs
from the Sunyaev and Zel'dovich result (see e.g. Rephaeli
et al. \cite{Rephaeli95}). This temperature dependence of the
S.Z. spectrum is particularly important in the submillimetre domain. 
Pointecouteau et al. (\cite{Pointecouteau98}) have shown that
it could be used to
determine the electron temperature from future space-borne sensitive 
submillimetre/millimetre S.Z. data, independently of any Xray measurement.

The comparison of the cluster's Xray flux with the amplitude
of the S.Z. thermal effect allows to derive the cluster's angular
distance and thus the Hubble constant. This method is fully
independent of the classical distance-scale determination.
Number of measurements of the decremental
part at millimetre or centimetre wavelengths have been successfully achieved from the 
ground and allowed determinations of peculiar velocities or $\Ho$
 (see for instance Birkinshaw et al. \cite{Birkinshaw91}, Jones et al.
\cite{Jones93}, Birkinshaw and Hughes \cite{Birkinshaw94},
Myers et al. \cite{Myers97}, and Holzapfel et al. \cite{Holzapfel97vpec},
\cite{Holzapfel97h0}).
The determination of the clusters peculiar velocity is performed
at wavelengths near 1.4 mm, where the thermal effect is close to zero.
At this wavelength the dust emission of galactic cirrus or extragalactic
starbursts can be significant and polutes the measurement. 
Only Infrared/submillimeter 
observations can raise this degeneracy. We report in this paper such
measurements for the cluster Abell 2163.

This is a well studied massive cluster located at $z = 0.201$, 
which has one of the hottest known intracluster gas temperatures 
($T_e = 12-15 keV$, Elbaz et al. \cite{Elbaz95} and
Markevitch et al. \cite{Markevitch96}). 
This makes it a very attractive candidate for 
measurements of the S.Z. effect.
High sensitivity measurements of the S.Z. effect on this cluster
have been obtained at 1.1, 1.4 and 2.1 mm 
using the SuZIE Caltech photometer on the CSO (
Holzapfel et al. \cite{Holzapfel97inst}).

\section{\label{observations} Observations}

\subsection{\label{sbmm} Submillimetre data}

The submillimetre data presented here were 
obtained with the PRONAOS telescope 
\footnote[1]{The PRONAOS telescope 
 is operated under the responsibility of the Centre National
 d'Etudes Spatiales (CNES), in collaboration
 with the laboratories of the authors} 
(Serra et al. \cite{Serra98})
during a 30-hour stratospheric flight from Fort Sumner (N.M.) on September the 22nd
and 23rd, 1996.
The 2-metre primary dish of the instrument is a 6-element segmented mirror
with active control of the shape to compensate for inflight thermal and gravity
changes. The SPM photometer (Lamarre et al. \cite{Lamarre94}) includes warm optics providing 
the sky modulation and internal regulated black-body calibrators. The detector system 
uses bolometers
cooled at 0.3 Kelvin with closed cycle $^3He$ coolers inside a liquid $^4He$ cryostat. 
There are 4 spectral bands at 170-240 $\mic$,
240-340 $\mic$, 340-540 $\mic$, 540-1050 $\mic$ with beams of 2, 2, 2.5, and 
3.5 arcmin (channels 1 to 4 respectively). 
The optical scheme uses dichroics so that a single direction on the sky 
is observed in the four channels simultaneously. 

\begin{figure}
\plotone{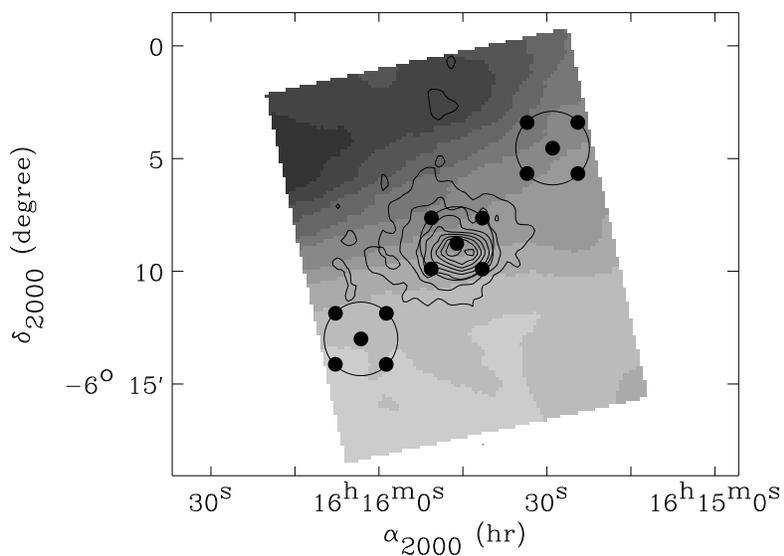}
\caption{Positions observed with PRONAOS overlaid on the 180 $\mic$
ISO-PHT map (greytones, min and max colors are respectively at 15 and
30 MJy/sr) and the X-ray ROSAT PSPC contours. The 
dots show the centers of the 5 observed positions (Cluster plus 
left and right beams at the same elevation), 
and the circles show the PRONAOS 3.5-arcmin beam FWHM at 630 $\mic$.
\label{fig1}}
\end{figure}

Our observation strategy includes three levels of beam modulation to subtract 
the instrumental systematic signals and their possible variations: 
i) beam switching at constant elevation, with a 6-arcmin amplitude, at a 
frequency of 19.5 Hz, 
ii) telescope nodding so that the source is alternately seen in the positive 
and negative beams, at a frequency of about 0.01 Hz, 
iii) repeating the observation on a blank field, 
with the same duration and observing mode. 
The blank field was selected in the neighborhood 
of the cluster, for minimum dust contamination in the IRAS 100$\tiretmic$
maps ($\alpha_{50}= 17h17m58s, \delta_{50} = 4d36'32"$).
Five positions are observed in the direction of the cluster: 
one at the nominal cluster 
centre (maximum X-ray brightness) and the other four at 1.6-arcmin offsets in elevation 
and cross-elevation from this centre 
(see Fig. \ref{fig1}). This strategy was chosen to cope with possible
miss-pointing of the telescope (which did not occur) and to provide
information on the cluster extension in case of a high signal-to-noise
detection. A2163 and the reference field were each observed
for a total duration of 54 minutes.
 
The sensitivities achieved in flight by the PRONAOS-SPM
system are 1.0, .58, .17 and .06 MJy/sr respectively in the four channels,
for S/N=1 and an observation time of 1 hour. The noise level in channels 1
and 2 is dominated by low frequency fluctuations likely to be due to residual 
atmospheric emissions. These fluctuations are still important
in channel 3, but can be efficiently removed by correlation with channel 1
and subtraction of the correlated component. The residual noise level
is then about 0.09 MJy/sr in channel 3, which, together 
with the noise in channel 4, is close to what is expected from 
photon noise. The calibration is performed in four steps: 
i) ground based calibration of the photometer on extended blackbodies 
at different temperatures, 
ii) ground based measurement of the bands' spectral responses, 
iii) inflight measurement of the instrument beam on the planet Saturn, 
and iv) inflight monitoring of the detector response on an internal 
modulated reference source
 (a detailed description of the noise analysis and
calibration procedures will be published in a forthcoming paper
by Pajot et al.). The resulting absolute calibration accuracy has
been estimated to be of order 10\%. Finally, for each analysed source
the calibration coefficients are re-computed in a self
consistent manner, integrating the bandpass over the best
estimate of the source's spectrum: e.g. dust plus S.Z. effect in the case
of A2163. 

The data processing includes: i) data deglitching and filtering, 
ii) demodulation from telescope nodding:
$(positive_{beam} - negative_{beam})/2$, iii) averaging.
Before averaging, channels 3 and 4 were cross-correlated with channel 1 to 
subtract any correlated noise component. The correlation ratios found are 
$F_3/F_1 \simeq 0.06 $ and $F_4/F_1 \simeq 0.03$ in Jansky units. These color 
ratios are much smaller than the cosmic dust emission colors, which 
will be the main contributor to the flux measured in channel 1: $F_3/F_1 \simeq 
0.43 $ and $F_4/F_1 \simeq 0.11$ for dust at 15 K 
with a spectral index of 2. However, as this decorrelation process
also implies some subtraction of the dust component from channels 3 and 4, 
the average data values were self-consistently corrected using the final
dust best fit spectrum ($T \simeq 15 K$). 

\subsection{\label{fir} Far infrared data}

\begin{figure}
\plotone{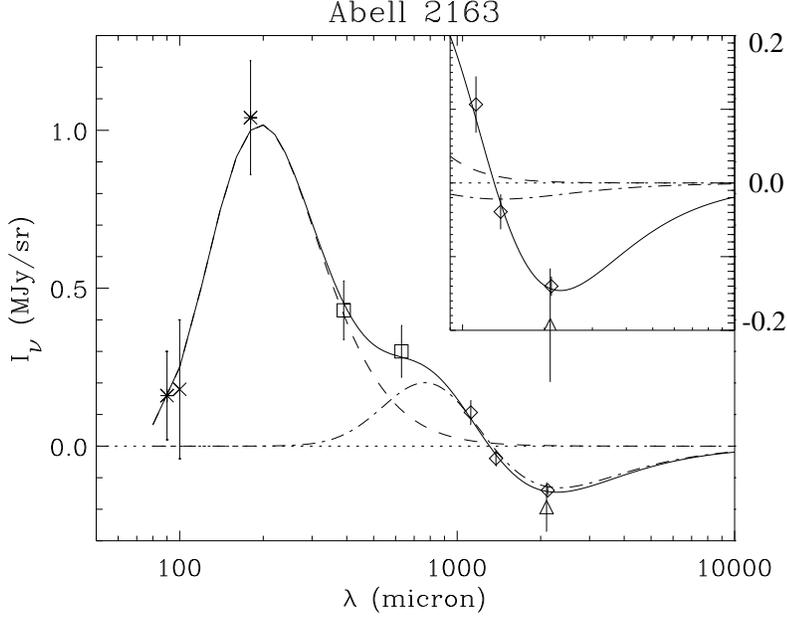}
\caption{Far infrared to millimetre brightness difference between  
the cluster of galaxies A2163 and its surroundings, using data 
from ISO-PHT ($\ast$), IRAS ($\times$), PRONAOS ($\Box$), 
SuZIE ($\Diamond$), Diabolo ($\Delta$). All values are homogeneous with the 
PRONAOS dilution (3.5-arcmin beam, 6-arcmin beamthrow). The solid line 
shows the best fit of a model including the contributions of: foreground 
dust (dash line), positive and negative parts of the 
thermal S.Z. effect (dash-dot line), and
kinetic S.Z. effect (dash-dot line in the insert). The parameters of the fit 
are: dust temperature $T_d = 14.8 \pm 1$ K, comptonization parameter 
in the direction of the cluster centre 
$Y_c = 3.42 (+0.41,-0.46) 10^{-4}$, cluster peculiar velocity 
$V_{pec} = 975 (+812,-971)$ km/s which can also be interpreted as a
negative CMB fluctuation of -119(+99,-118) $\mu K$.
\label{fig2}}
\end{figure}

Maps at 90 $\mic$ and 180 $\mic$ were obtained with the PHOT instrument on board
the ISO satellite, (open time program: GSERRA:SZCLUST). The observation
mode was a PHOT-32 8x8 raster map with steps of 92 arcsecond and
integration times of 20 and 64 seconds per pixel (pixel sizes are
43 and 90 arcsec respectively for the 90$\tiretmic$ and 180$\tiretmic$ filters). 
The data reduction was performed with the PIA-7.01 software. In order to get 
consistent data allowing us to derive the 
dust emission, the PRONAOS observing sequence was simulated
with its parameters -beam switching and beam size- on the ISO maps 
(see Fig. \ref{fig1}).
The results of these simulated measurements are reported 
in Table \ref{table1} with the 
corresponding error bars (see also Fig. \ref{fig2}). 
The error bars have been derived from the rms fluctuation 
taken on a map of residuals. This map was obtained by subtraction 
of the large scale cirrus pattern fitted as a 5 degree 2D
polynomial. The error bar obtained in this manner is always 
larger than the PHOT absolute calibration uncertainty (10 
to 20 \%). 
 
The same operation was performed on the IRAS
ISSA map at 100 $ \mic$. Although the spatial modulation for PRONAOS 
and SuZIE is not exactly the same, the simulations using the 180$\tiretmic$ 
ISO-PHT data show that the dust signal is the same within error bars. 
Thus a single dust signal is used in the interpretation of the data.

\begin{deluxetable}{rlll}
\footnotesize
\tablecaption{Numerical values for the measurements gathered
in Fig. \ref{fig2}. \label{table1}}
\tablewidth{0pt}
\tablehead{
\colhead{$\lambda$} & \colhead{Instrument}   &
\colhead{$Y_c$\tablenotemark{a}} & 
\colhead{MJy/sr\tablenotemark{b}}\\
\colhead{$(\mic)$} &  & $(\dix^{-4})$ & 
} 
\startdata
90 & ISO-PHT & & 0.16 [0.14] \nl
100 & IRAS & &  0.18 [0.22] \nl
180 & ISO-PHT  & & 1.04 [0.18] \nl
390 & PRONAOS  & & 0.43 [0.09] \nl
630 & PRONAOS  & & 0.30 [0.08] \nl
1120 & SuZIE &   4.05 [1.45] \tablenotemark{c} & 0.107 [0.038] \nl
1380 & SuZIE &   9.9 [6.0] \tablenotemark{c} & -0.039 [0.024] \nl
2120 & SuZIE &  3.73 [0.35] \tablenotemark{c} & -0.140 [0.013] \nl
2100 & Diabolo &   5.5 [2.2] \tablenotemark{d} & -0.20 [0.08] \nl
\enddata

\tablenotetext{a}{Central comptonization parameter taken from the litterature
  (assuming the same King profile as in this paper).}
\tablenotetext{b}{Estimated signal for a 3.5' beam and 6' modulation
  amplitude (square brackets give $1\sigma$ errors).} 
\tablenotetext{c}{Holzapfel et al. \cite{Holzapfel97vpec} and \cite{Holzapfel97h0}}
\tablenotetext{d}{D\'esert et al. \cite{Desert98}}
\end{deluxetable}

\section{\label{discussion}Results and discussion}

We show in Fig. \ref{fig2} our far-infrared and submillimetre
measurements together with the existing millimetre data from 
Holzapfel et al. (\cite{Holzapfel97vpec}) at 1.1, 1.4 and 2.1 mm 
(SuZIE experiment on the CSO telescope: 
1.9 arcmin beam, 4.6 arcmin beamthrow) and 
from D\'esert et al. (\cite{Desert98})
at 2.1 mm (Diabolo experiment on the IRAM 30 metre telescope: 
0.5 arcmin beam, 3 arcmin beamthrow). The numerical values
with $1 \sigma$ error bars are reported in Table \ref{table1}.
The Diabolo and SuZIE data have been corrected for the different beams and modulation
amplitudes to compare with PRONAOS data, 
assuming a King profile for the intracluster gas density:

\begin{equation}
n_{gas}(r) = n_0 \left[ 1+ \left( \frac{r}{r_c} \right)^2
\right] ^{-3\beta/2}
\label{eq:kingprof}
\end{equation}

where $r_c$, the cluster core radius, corresponds to a projected angle 
$\theta_c = 1.2$ arcmin and $\beta = 0.62$ from Elbaz et al. (\cite{Elbaz95}).
The dilution factors for PRONAOS and Diabolo are respectively 
0.38 and 0.55 in terms of the
ratio of the beam averaged signal to the central cluster value. For PRONAOS
this also takes into account averaging over the offset positions. For SuZIE
we use the values derived from single band fits by Holzapfel et al.
(\cite{Holzapfel97vpec}),
which use the same cluster model as we do.

Different components appear in this spectrum: the dust shows up 
at shorter wavelengths and is measured at 90 $ \mic$ (ISO), 
100 $ \mic$ (IRAS), 180 $ \mic$ (ISO) and 390 $ \mic$ (PRONAOS). 
The S.Z. thermal effect has its maximum positive
and negative peaks respectively at submillimetre and millimetre wavelengths.
The positive part is measured at 630 $ \mic$ (PRONAOS) and 1.1 mm (SuZIE),
whereas the negative part is measured at 2.1 mm (SuZIE and Diabolo). 
Finally the Kinetic S.Z. effect (or CMB primordial temperature fluctuation,
see below) is dominant
in the SuZIE 1.4 mm band, where the S.Z. thermal effect is close to zero.\\

We have simultaneously fit the ISO, 
PRONAOS, Diabolo and SuZIE data with a S.Z.
plus dust spectrum having four free parameters: $Y_c$ the comptonization parameter
in the direction of the cluster center, $V_p$ the peculiar velocity of the
cluster, $F_d(180)$ the level of dust emission at 180 $\mic$, and 
$T_d$ the dust temperature (The dust emissivity index being fixed to $n_d = 2$ 
from Boulanger et al. \cite{Boulanger96}). We are not able to distinguish the S.Z. kinetic 
effect from a thermal fluctuation of the
CMB itself since they have the same spectrum. 
In the following we will quote both the velocity
value and the peak $\Delta T$ value for a CMB fluctuation 
that would have the same angular distribution and flux as the cluster S.Z.
kinetic effect.
For the S.Z. effect we assume 
an intracluster gas temperature of 13 keV and use an exact 
relativistic thermal S.Z. spectrum from Pointecouteau et al. (\cite{Pointecouteau98}).
The best fit values for the four parameters using the whole data set, 
and 68\% confidence intervals, are: 
$Yc = 3.42 (+0.41,-0.46) 10^{-4}$, $V_p = 975 (+812,-971)$ km/s 
(cluster moving away from us, $\Delta T_{CMB} = -119 (+99,-118) \mu K$), 
$F_d(180) = 1.00 (+0.09,-0.16)$ MJy/sr and $T_d = 14.8 \pm 1$ K. 
The error bar on each parameter has been calculated from a likelihood analysis
after integration over the other parameters: $L = exp(-\chi^2/2) $.
It was not necessary to assume a prior probability for any of the four
parameters.

First of all, this data set provides three measurements of the
S.Z. effect in A2163 made with significantly different beam sizes, 
3.5, 1.9 and 0.5 arcmin respectively for PRONAOS, SuZIE and
Diabolo. Within the error bars, the three measurements give 
the same value for the comptonization
parameter in the direction of the cluster center. 
This validates the King profile that has been derived
from the X-ray analysis for the intracluster gas distribution, for
a cluster radius ranging from 0.4 to 5 times the core radii.

\begin{figure}
\plotone{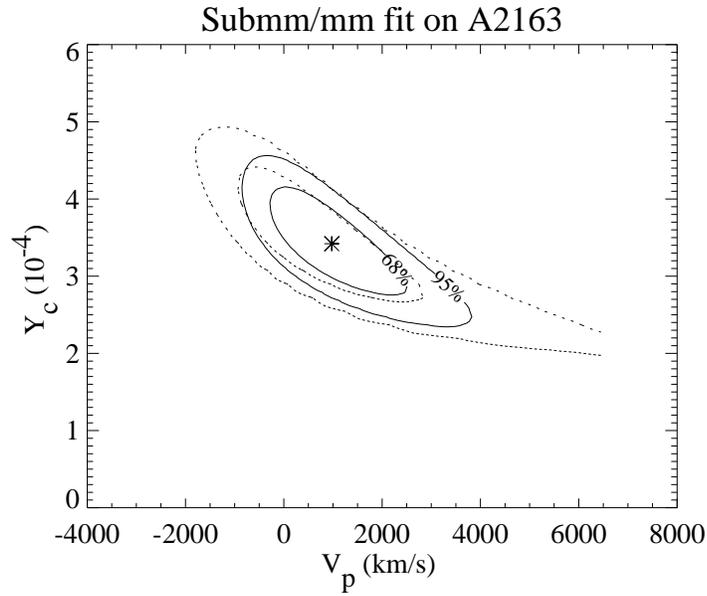}
\caption{68\% and 95\% confidence limits for the determination of the cluster
peculiar velocity and central comptonization parameter from the far-infrared,
submillimetre and millimetre data. The dashed line shows the degraded limits
if far-infrared and submillimetre data are ignored, allowing the dust contribution to 
vary in a plausible range (see text).
\label{fig3}}
\end{figure}

Our values for the compton parameter and cluster peculiar velocity
can be compared with the values derived in Holzapfel et al. (\cite{Holzapfel97vpec})
from the analysis of the SuZIE data at 1.1, 1.4 and 2.1 mm: 
$Y_c = 3.73 (+0.47,-0.61) 10^{-4}$ and $V_p = 490 (+910,-730) km/s$.
However in their analysis the authors assumed a zero dust emission level, 
which is not realistic given the far infrared measurements. The ISO map
in Fig. \ref{fig1} clearly shows that this cluster, which is at
medium galactic latitude, lies behind
a significant layer of galactic dust (about 25 MJy/sr at 180 $\mic$).
We show in Fig. \ref{fig3} that 
the millimetre data itself does not allow a precise
determination of the cluster peculiar velocity in the absence of any 
submillimetre measurement constraining 
the dust contribution. This figure displays the
68\% and 95\% confidence contours on
$V_p$ and $Y_c$, including (and not including) the far
infrared and submillimetre data in the fit: full lines 
(dashed lines). In the second case (dashed line) 
the parameters of the dust component have been varied in the
following range:
$-2 MJy/sr < F_d(180 \mic) < 2 MJy/sr$, $10 K < T_d < 20K$. 
This is a plausible range considering the average depth of the
cirrus cloud in the direction 
of A2163 : $\NH = 1.6 \dix^{21} \cmdeux$ from 
the Xray data (Elbaz et al. \cite{Elbaz95}), 
or $\NH = 2.1 \dix^{21} \cmdeux$ from 21 cm surveys
(Dickey and Lockman \cite{Dickey90}). We use  
the infrared to $N_H$ colors from 
Boulanger et al. (\cite{Boulanger96}): $1.26 \dix^{-20} MJy/sr/cm^{-2}$
at $140 \mic$.

Although it remains within the previous error bars, the velocity change 
induced by taking into account the dust contribution implies
a slight decrease in $Y_c$. This is because the kinetic S.Z.
effect contributes about 10\% of the measured decrement at 2.1 mm, 
the wavelength where the thermal S.Z. effect is mostly constrained
by the data. The combined
data set (submillimetre plus millimetre) only slightly 
improves the precision of the comptonization parameter
since the accuracy of the PRONAOS measurement of the positive
S.Z. effect (S/N = 3 to 4) is much less than that of the SuZIE measurement
at 2.1 mm (S/N = 10). 

Concerning the derivation of $\Ho$ by comparison
of the comptonisation parameter with X-ray fluxes, this scales as 
$F_X/Y^2$. Thus, the $\Ho(q_o=1/2)$ value derived using the same
isothermal model for the cluster gas distribution as 
Holzapfel et al. (\cite{Holzapfel97vpec}) increases from $60 (+40,-23)$
to $71 (+47,-27) km/s/Mpc$ with our determination of $Y_c$.\\

To conclude, the combined data set presented here (Fig. \ref{fig2}) 
demonstrates the need for submillimetre data to correctly
interpret S.Z. measurements and more generally millimetric CMB data. 
At this stage we are not able to assess the origin
of the submillimetre dust emission towards A2163: 
i) residual unbalanced galactic dust as
assumed above, 
ii) background starburst galaxies which can possibly be magnified by a 
gravitational lensing effect
or iii) intracluster dust as measured by Stickel et al. (\cite{Stickel98})
toward the Coma cluster. This last hypothesis was 
carefully examined. Data from ISO-PHT show that the measured flux 
depends on the orientation of the simulated beam switching, and 
that the mean of all possible orientations (1.5 $ \sigma$ detection) 
does not provide a confirmation of a brightness excess 
towards the cluster. Moreover, given the depth of the foreground
cirrus cloud (see above) the 
hypothesis of contamination by galactic dust seems very likely. 

The case of A2163, a massive 
cluster at medium galactic latitude,
actually provides a good illustration of the problems that
will be faced when interpreting data for more standard clusters at 
higher galactic latitudes ($Y_c
\simeq 10^{-5}$, $F_d(180 \mic) \simeq 5 $ MJy/sr) 
with the future very sensitive space borne 
instruments such as
Planck (Bersanelli et al. \cite{Bersanelli96}). Submillimetre capabilities, 
as provided by
Planck and FIRST, will be required to
derive meaningful peculiar velocities, and precisions below 10\% on the
comptonisation parameter. However, concerning the peculiar velocity,
the fundamental limitation will remain the CMB because it has
the same spectrum as the S.Z. kinetic effect  (see e.g. Haehnelt
and Tegmark \cite{Haehnelt96}
and Aghanim et al. \cite{Aghanim97}).
In a similar way, the analysis of the primordial CMB anisotropies themselves
at high angular resolution ($\theta > 5'$) from the Planck 217
GHz channel, will also suffer from the dust contamination and require
the help of the submillimetre channels.

\acknowledgments

We are very indebted to the CNES team 
led by F. Buisson who were in charge of the development and operations of the
PRONAOS gondola and telescope. The professionalism of the
NSBF team led by D. Ball was very much appreciated. We are
also very grateful to the CNRS technical team led by G. Guyot, 
in charge of the focal plane instrument SPM. We thank W.L. Holzapfel and 
D.L. Clements for their help and knowledgeable comments.

\end{document}